\documentclass[preprint]{ptephy_v1}

\usepackage{hyperref}

\usepackage{threeparttable}
\usepackage{multirow}
\usepackage[separate-uncertainty=false]{siunitx}
\usepackage{subfig}

\begin{document}

\title{
  Response of germanium detectors for high-energy $\gamma$-rays
  by $^{27}$Al($p,\gamma$)$^{28}$Si at $E_p=992$~keV}


\author[1,*]{Rurie~Mizuno}
\affil{Department of Physics, the University of Tokyo, 7-3-1 Hongo, Bunkyo-ku, Tokyo 113-0033, Japan}

\author[1,2]{Megumi~Niikura}
\author[2]{Tokihiro~Ikeda}
\author[2]{Teiichiro~Matsuzaki}
\author[2]{Shintaro~Go}
\affil{RIKEN Nishina Center for Accelerator-Based Science, 2-1 Hirosawa, Wako, Saitama 351-0198, Japan}
\author[1,3]{Takeshi~Y.~Saito}
\author[3]{Shin'ichiro~Michimasa}
\affil{Center for Nuclear Study, the University of Tokyo, 2-1 Hirosawa, Wako, Saitama 351-0198, Japan \email{mizuno@nex.phys.s.u-tokyo.ac.jp}}
\author[1,2]{Hiroyoshi~Sakurai}


\begin{abstract}%
The performance of germanium detectors for high-energy $\gamma$-rays was evaluated using 
a 992-keV resonance in the $^{27}$Al($p,\gamma$)$^{28}$Si reaction.
The measurement was conducted at the RIKEN tandem accelerator.
The energy of the excited state from the resonance was evaluated as 12540.7(2) keV. 
Using newly evaluated excitation energy, an energy calibration function and a photo-peak efficiency of Ge detectors up to 10.8-MeV photon were deduced. The energy accuracy is achieved at 0.3 keV for the overall energy region. 
This reaction provides reliable energy and efficiency standards for high energy $\gamma$ rays.
\end{abstract}

\subjectindex{H15, D50}

\maketitle

The muonic atom is a bound system of a muon and a nucleus, and
the X-ray spectroscopy of the muonic atom is used 
to determine the charge radius of the nucleus~\cite{Fricke1995,Angeli2013,Powers1979,Bergem1988,Engfer1974,Antognini2020,Saito2022}.
The energy of muonic X rays of light elements is about a few tens keV, e.g. 75 keV for carbon K$_\alpha$ line, 
while the X-ray energy of heavy elements, such as $^{208}$Pb, is about 6 MeV
~\cite{Fricke1995}.
We have been developing a photon detection system with a wide dynamic range
using high-purity germanium (Ge) detectors 
for X-ray spectroscopy of the muonic atom.
The evaluation of the energy accuracy is crucial 
for the determination of the nuclear charge radius from the muonic X-ray energies. 
In the case of $^{208}$Pb, for example, 0.3-keV accuracy of the X-ray energy corresponds to 0.02\% accuracy of the charge radius, and is comparable to the uncertainty of the theoretical model~\cite{Bergem1988}.

For muonic X-ray spectroscopy, the detector responses, such as detection efficiency 
and resolution, 
should be evaluated 
in wide energy region. 
While standard $\gamma$-ray sources are used for the evaluation of detector performance
in low-energy regions below 1.5 MeV, 
available $\gamma$-ray sources are limited for high-energy regions.
We adopted a $\gamma$-ray measurement of a 992-keV resonance in the $^{27}$Al($p,\gamma$)$^{28}$Si reaction for evaluation of the detector responses for high energy photons.
This reaction emits several $\gamma$ rays over a wide energy range
from 1.5 to 10.8 MeV.
~\cite{Antilla1977-ad, Endt1990-xs}.

\begin{figure}[!h]
  \centering\includegraphics[width=5in]{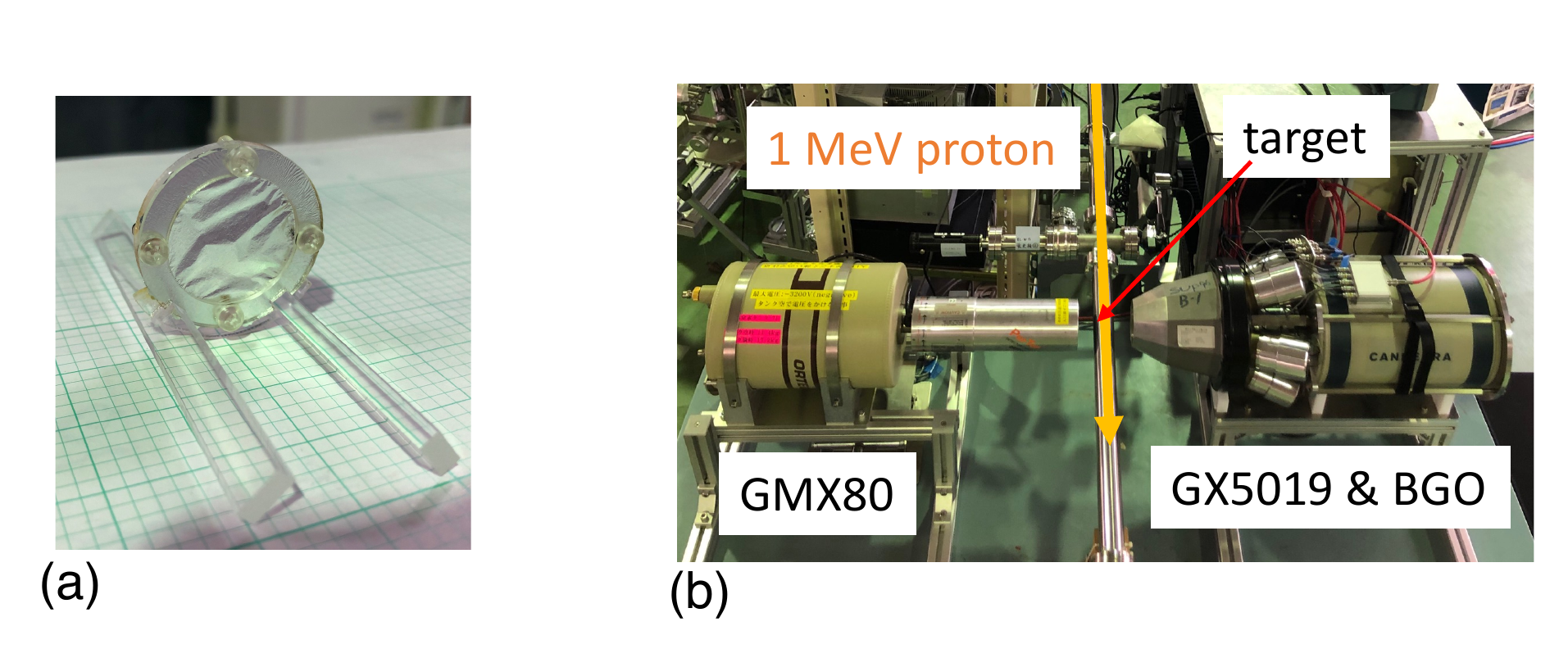}
  \caption{Photographs of the experimental setup. 
  (a) shows the aluminum target. The 0.8-$\mu$m aluminum foil was supported by an acrylic frame.
  (b) shows the overall setup.
  A yellow arrow represents the proton beam. 
  The target was placed in the vacuum (better than 1$\times10^{-3}$ Pa) duct.
  Two Ge detectors were installed 
   at an angle of 90 degrees respecting the beam axis.
  GX5019 was surrounded by the BGO crystals. BGO crystals are used as Compton suppressors, and their performance will be discussed in the forthcoming paper.}
  \label{fig:pelletron_setup}
\end{figure}

The experiment was performed at the RIKEN tandem accelerator (Pelletron 5SDH-2, 1.7 MV max).
The experimental setup is shown in Fig.~\ref{fig:pelletron_setup}. 
A 0.8-$\mu$m thick aluminum target (20 mm$^\phi$) was irradiated with the proton beam at 1~MeV in the beam duct (1-mm thick, made of stainless). 
The proton beam intensity was approximately 300 nA, and the measurement time was 6.5 hours. 
Two large volume Ge detectors, GMX80 (Ortec) and GX5019 (Canberra), were used for $\gamma$-ray detection. 
GMX80 is an n-type 80\% coaxial detector (with a diameter of 74.8 mm$^\phi$ and length of 76.4 mm), and GX5019 is a p-type 50\% coaxial detector (with a diameter of 62 mm$^\phi$ and length of 75.1 mm).
The distance from the target to the detector surface was 5 cm (GMX80) and 10~cm (GX5019), respectively.
Signals from the detectors were acquired by a waveform digitizer (Caen V1730B, sampling rate of 500MHz and resolution of 14bit).
The maximum energy of the input dynamic range of the digitizer was set to about 20 MeV.


The target thickness was decided to be 0.8 $\mu$m, and the beam energy is chosen to be 1.00 MeV. 
The target thickness and the beam energy induce
the 992-keV resonance without populating the nearest resonances of $^{27}$Al+p at 937 and 1025 keV. 
The 937 and 1025-keV resonances must be avoided 
because the $\gamma$-ray intensity is known only for the 992-keV resonance.
By considering the energy fluctuation of the incident beam at 1 MeV ($\pm$15 keV measured from the terminal voltage monitor of the tandem accelerator) and the energy struggling in the target ($\pm$4 keV), the energy loss of 1 MeV proton in the target has to be less than 44 keV to avoid the 937-keV resonance. 
The energy loss of the 1.00 MeV proton beam in the 0.8-$\mu$m Al target is 36 keV, which satisfies the condition. 
The target thickness was measured with the energy loss of $\alpha$-rays from triple-$\alpha$ standard sources ($^{148}$Gd, $^{241}$Am, and $^{244}$Cm) prior to the experiment.
The triple-$\alpha$ sources, the target, and a Si pin-photo diode were set in a vacuum chamber. 
From the energy loss in the target measured with the Si detector, 
the target thickness was deduced to be 0.764(16) $\mu$m.

The performance of the Ge detectors below 1.5 MeV is measured with the standard $\gamma$-ray 
source of $^{152}$Eu.
The non-linearities of the digitizer and the pre-amplifier were measured with a pulse generator (Ortec 448) in the full input dynamic range, 
and corrected with the 4th-order polynomial function.

\begin{figure}[!h]
  \centering
  \includegraphics[width=14cm]{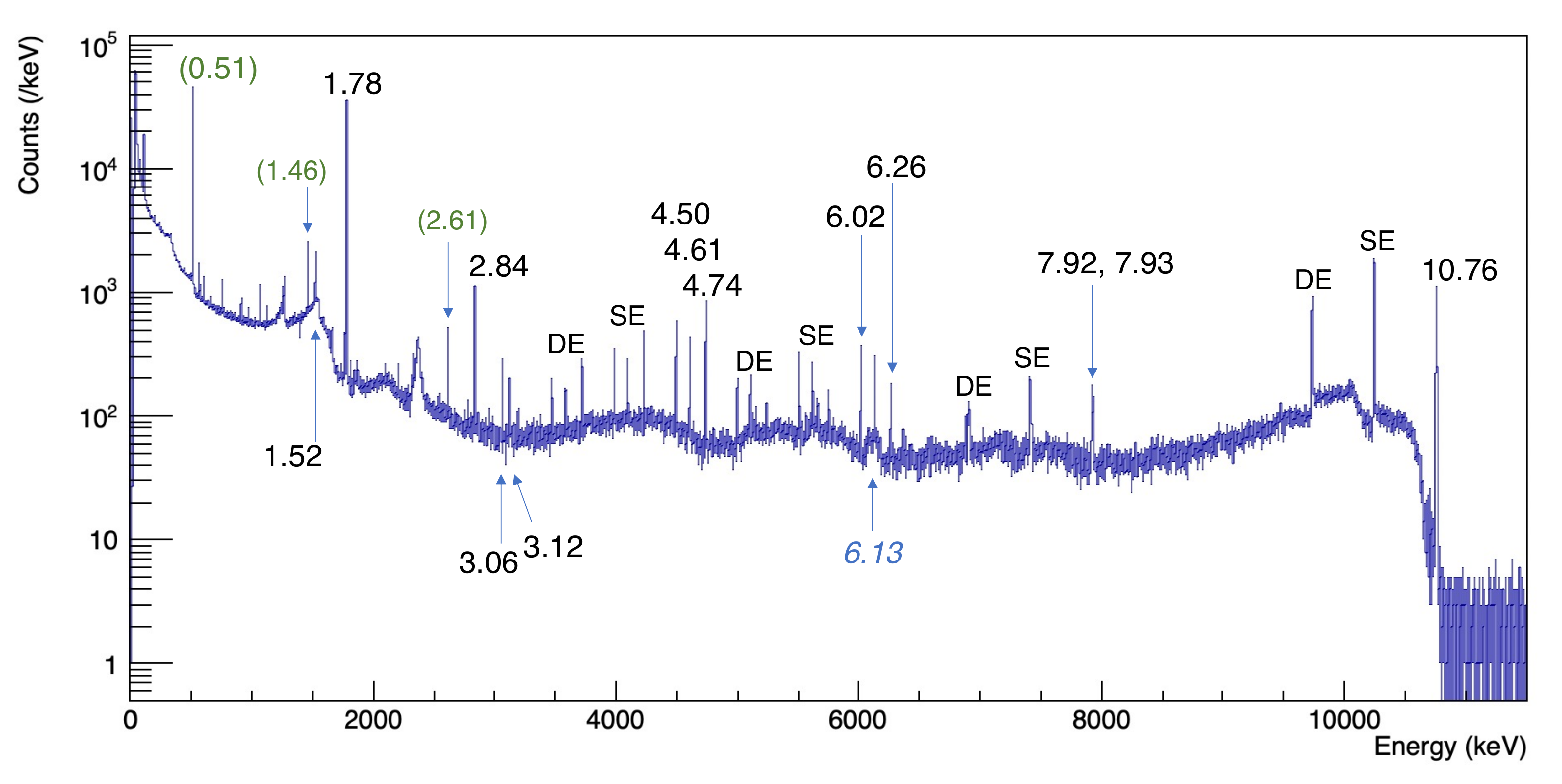}
  \caption{$\gamma$-ray energy spectrum measured by GX5019. 
  The numbers written in the figure show the $\gamma$-ray energies of the peaks in MeV. 
  Thirteen $\gamma$-ray peaks from the $^{27}$Al($p,\gamma$)$^{28}$Si reaction are observed. 
  In the spectrum, SE (DE) means single (double) escape peaks corresponding to the $\gamma$-rays.
  $\gamma$-ray peaks at 0.51, 1.46, and 2.61 MeV are natural backgrounds.
  }
  \label{fig:spectrum}
\end{figure}

The $\gamma$-ray energy spectrum acquired with GX5019 is displayed in Fig.~\ref{fig:spectrum}. 
The spectrum shows 13 $\gamma$-ray peaks from the $^{27}$Al($p,\gamma$)$^{28}$Si reaction in 1.5--10.8 MeV, as well as 
their single escape peaks (SE) and double escape peaks (DE). 
The natural background peaks, 510.999 keV from electron annihilation, 1460.820 keV from $^{40}$Ar, and 2614.511 keV from $^{208}$Pb were found in the spectrum.
All these energy peaks have more than 1000 counts, and were used for the energy calibration.
A $\gamma$-ray peak at 6128.63 keV originated from $^{16}$O was also found in the spectrum. The 6128.63-keV peak was not used for the calibration because the peak is a background which comes from the surrounding materials and depends on the measurement conditions.
The peaks have been fitted by a Gaussian function with a background term, and the Ge detectors' photo-peak efficiencies and energy calibration functions are deduced in a wide energy range below 10.8 MeV.



\begin{figure}[!h]
  \centering
  \includegraphics[width=8.5cm]{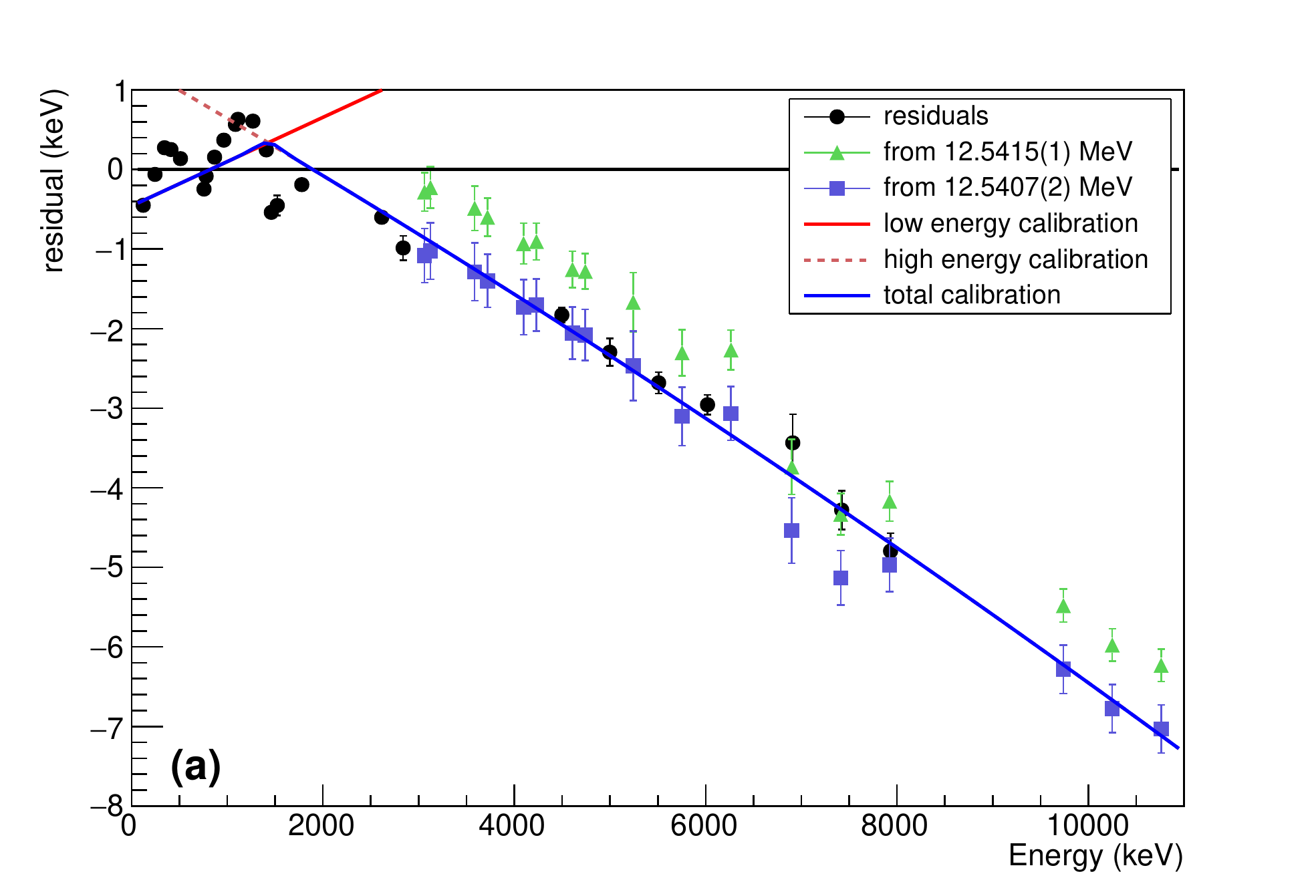}
  \includegraphics[width=8.5cm]{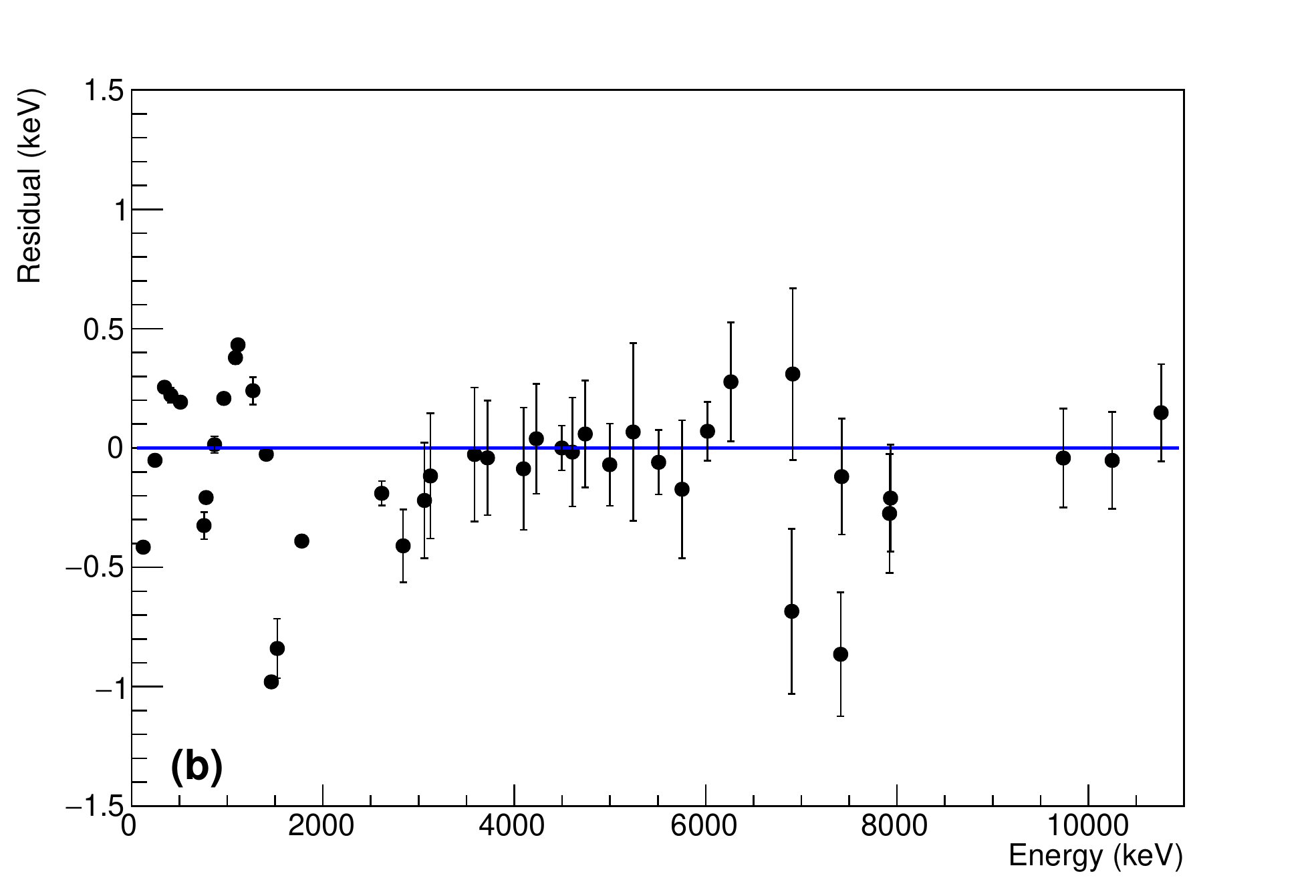}
  \caption{The residuals from calibration functions of (a) linear function, (b) Eq.~\ref{eq:Ecalib} for each peak taken with GX5019. The green triangle points in (a) represent $\gamma$-ray peaks from the state at 12.54 MeV, and the black circle points show the other $\gamma$-ray transitions. The blue square points represent the adopted residuals from the state at 12.54 MeV.}
  \label{fig:residual}
\end{figure}

Table~\ref{tab:energy_list} shows the final values of the reference $\gamma$-ray energies and relative intensities of 
the $^{27}$Al($p,\gamma$)$^{28}$Si resonance reaction at 992 keV used in the high energy $\gamma$-ray calibration.
High energy calibration method using the $^{27}$Al+p reaction was originally proposed by Antilla et al~\cite{Antilla1977-ad}, and
some $\gamma$-ray energies have been updated by the NNDC (National Nuclear Data Center) compilation~\cite{NNDC}.
At first, the energy calibration using the linear function was conducted, and the residual energies from the function deduced with GX5019 are shown in 
Fig.~\ref{fig:residual}(a). 
The green triangle points correspond to $\gamma$-ray transitions from the excited state at 12.54 MeV, and the black circle points show other transitions, $\gamma$-rays from the standard source and background $\gamma$-rays. 
As the $\gamma$-ray energies from the state at 12.54 MeV are not provided in NNDC, the energies of the $\gamma$-ray transitions are calculated from the excitation energy of initial and final states ($E_i$ and $E_f$) with considering the recoil effect.
The 12.5415(1) MeV was used as the excitation energy of the initial state which is taken from NNDC~\cite{NNDC}.
From the figure, the green points systematically deviate from the other points.
Therefore, the excitation energy of 12.5407(2) MeV was deduced by minimizing the sum of the square of all the $\gamma$-ray energy residuals between 2 and 8 MeV. 
Blue square points in Fig.~\ref{fig:residual}(a) are the residuals of the $\gamma$-ray peaks using the adopted excitation energy.
In Table~\ref{tab:energy_list}, the adopted and re-evaluated levels and $\gamma$-ray energies are summarised.
The relative intensities in Table~\ref{tab:energy_list} are taken from Antilla et al~\cite{Antilla1977-ad} for the efficiency calibration.

\begin{table}[!h]
  \begin{center}
    \centering
    \caption{The gamma-ray energies and intensities of $^{27}$Al+p resonance reaction at 992 keV.
    The values in the column of excitation energies of the initial ($E_i$) and final ($E_f$) states, and the energies of $\gamma$-ray transitions ($E_\gamma$) are taken from NNDC~\cite{NNDC} otherwise subscripts.
    The values in a column $I_\gamma$ are relative intensities of each $\gamma$-rays taken from Antilla et al~\cite{Antilla1977-ad}.}
    \label{tab:energy_list}
    \begin{tabular}[width=1.0\textwidth]{S[table-format=5.4]S[table-format=4.4]S[table-format=5.6]S}\hline\hline
      {$E_{i}$} & {$E_{f}$} & {$E_\gamma$ (keV)} & {$I_\gamma$ (\%)} \\ \hline
      7799.01(9)     & 6276.20(7)  & 1522.76(12)      & 2.8(2)\\
      1779.03(1)     & 0           & 1778.969(11)     & 94.8(15)\\
      4617.86(4)     & 1779.03(1)  & 2838.29(15)      & 5.5(4)\\
      12540.7(2)$^a$ & 9479.49(11) & 3061.13(23)$^b$  & 1.15(11)\\
      12540.7(2)$^a$ & 9417.17(14) & 3123.44(24)$^b$  & 0.70(7)\\
      6276.20(7)     & 1779.03(1)  & 4496.78(8)       & 4.8(3)\\
      12540.7(2)$^a$ & 7933.45(10) & 4606.94(22)$^b$  & 4.5(4)\\
      12540.7(2)$^a$ & 7799.01(9)  & 4741.36(22)$^b$  & 8.8(5)\\
      7799.01(9)     & 1779.03(1)  & 6019.28(10)      & 6.0(5)\\
      12540.7(2)$^a$ & 6276.20(7)  & 6263.85(21)$^b$  & 2.1(2)\\
      12540.7(2)$^a$ & 4617.86(4)  & 7921.74(20)$^b$  & 4.3(4)\\
      7933.45(10)    & 0           & 7932.24(11)      & 3.7(4)\\
      12540.7(2)$^a$ & 1779.03(1)  & 10759.55(20)$^b$ & 76.6(15) \\\hline\hline
      \end{tabular}
    \begin{tablenotes}
      \item[*] $^a$ Deduced by the present work. See text.
      \item[]  $^b$ Calculated values using $E_i$ and $E_f$ with recoil correction.
    \end{tablenotes}
  \end{center}
  \end{table}

From Fig.~\ref{fig:residual}(a), the energy of the high-energy region is different from the extrapolation from the calibration function of the low-energy region.
As the Ge crystal has a different trend in the linearity between below and above around 1.5 MeV, 
the following phenomenological calibration function is used,  
\begin{align}
  \label{eq:Ecalib}
  E ={}& f_1(N_\textrm{ch})\left[\frac{1}{2}\mbox{Erfc}\left(\frac{N_\textrm{ch}-p_5}{p_6}\right)\right]
  +f_2(N_\textrm{ch})\left[\frac{1}{2}\mbox{Erf}\left(\frac{N_\textrm{ch}-p_5}{p_6}\right)+\frac{1}{2}\right],\\ 
  \label{eq:Ecalib_f1}
  &f_1(N_\textrm{ch})=p_0+p_1 N_\textrm{ch},\\ 
  \label{eq:Ecalib_f2}
  &f_2(N_\textrm{ch})=p_2+p_3 N_\textrm{ch}+p_4 N_\textrm{ch}^2, 
\end{align}
which corresponds to the combination of a low energy calibration ($f_1$) and a high energy calibration ($f_2$) using the error function (Erf) and the complementary error function (Erfc).
The $p_{0-4}$ and $p_6$ are fitting parameters, $E$ is the energy, and $N_\textrm{ch}$ is a channel of the pulse height deduced from the digitized waveform. 
$p_5$ satisfies as $f_1(p_5)=f_2(p_5)$. 
The red and bashed light red lines in the Fig.~\ref{fig:residual}(a) show $f_1$ and $f_2$, respectively 
and the blue line shows the connected function expressed by Eq.~\ref{eq:Ecalib}.
Figure~\ref{fig:residual}(b) shows the residuals for each $\gamma$-ray peak from the calibration function using Eq.~\ref{eq:Ecalib}.
The accuracy of the energy is deduced as 0.3 keV in a wide dynamic range below 10 MeV after the non-linearity correction. 
This uncertainty may be caused by the higher-order component of the non-linearity, which can not be corrected by source calibration. 
0.3 keV accuracy satisfies the requirement of the muonic X-ray spectroscopy.

\begin{figure}[!h]
  \centering
  \includegraphics[width=9cm]{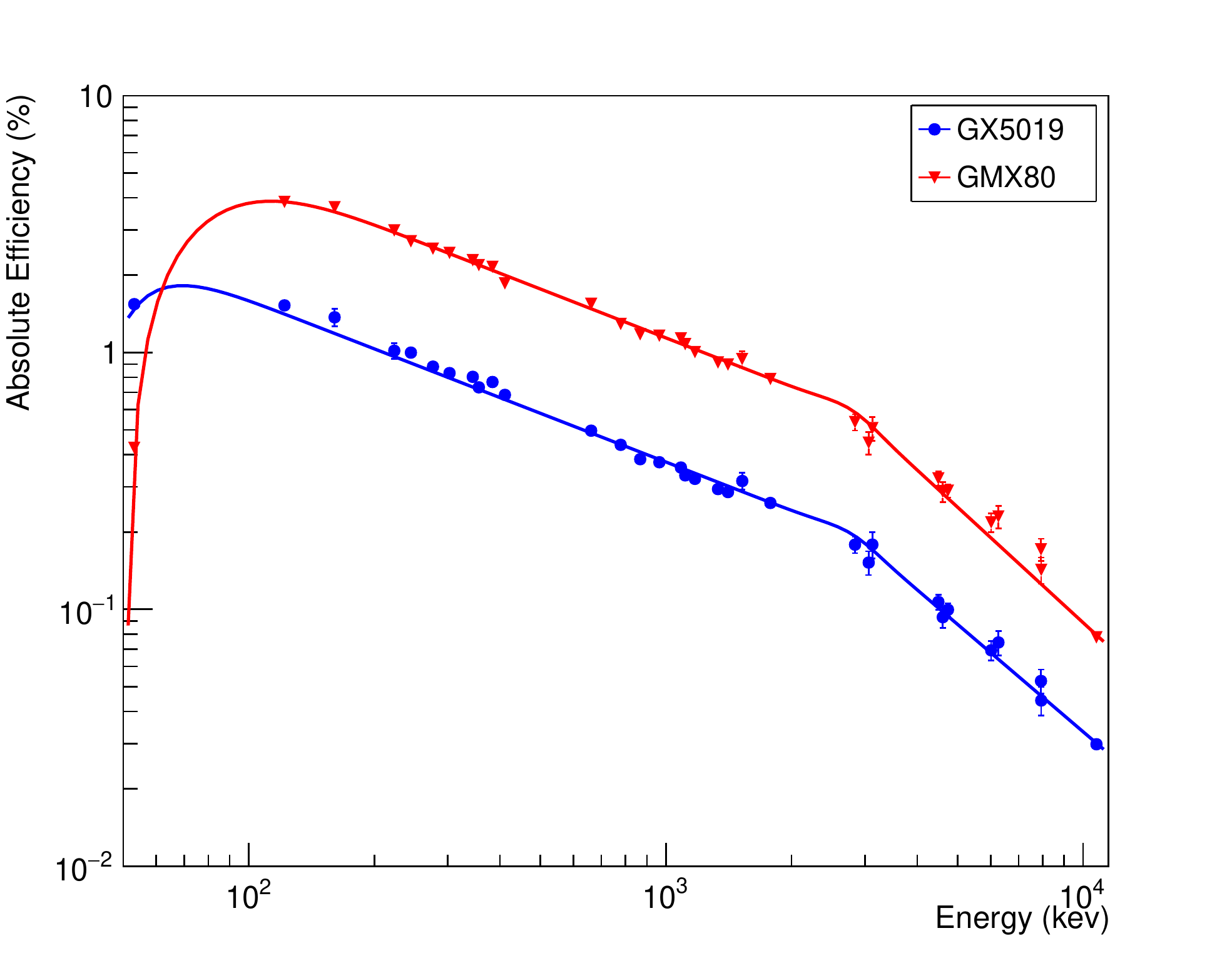}
  \caption{The absolute photo peak efficiencies of GX5019 and GMX80.}
  \label{fig:efficiency}
\end{figure}

The relative photo peak efficiencies of the Ge detectors below 10.8 MeV are also deduced
 using the intensity of all peaks in Table~\ref{tab:energy_list} and the $^{152}$Eu source.
Since the proton beam intensity was not measured in the present experiment, only the relative photo peak efficiencies from 1.5 to 10.8 MeV are obtained.
The absolute photo peak efficiency normalized to low energy region taken with the standard source is shown in Fig.~\ref{fig:efficiency}. 
From the figure, the efficiency curve in the high-energy region differs from the extrapolation of the low-energy region.
The solid lines shown in Fig.~\ref{fig:efficiency} are fitting functions; 
\begin{align}
    \label{eq:eff}
    \epsilon ={}& g_1(E)\left[\frac{1}{2}\mbox{Erfc}\left(\frac{E-q_5}{q_6}\right)\right]
    +g_2(E)\left[\frac{1}{2}\mbox{Erf}\left(\frac{E-q_5}{q_6}\right)+\frac{1}{2}\right],\\ 
    \label{eq:eff_g1}
    &g_1(E)=q_0(E^{-q_1}-q_3 \textrm{e}^{-q_4 E}),\\
    \label{eq:eff_g2}
    &g_2(E)=q_7 E^{-q_8}, 
  \end{align}
where $\epsilon$ is the efficiency, the $q_{0-4}$ and $q_{6-8}$ are fitting parameters, and 
$q_5$ satisfies as $g_1(q_5)=g_2(q_5)$.
Equation~\ref{eq:eff} well reproduces the photo peak efficiencies in a wide dynamic range.
Similar kinks at around 2-3 MeV of the photo-peak efficiency are previously reported~\cite{Singh1971,McCallum1975,Molnar2002,Elekes2003}, and explained with a model including pair production and multiple processes~\cite{Hajnal1974}.

In conclusion, the performance of the germanium detectors for high-energy $\gamma$-rays was evaluated using the 992-keV resonance in the $^{27}$Al($p,\gamma$)$^{28}$Si reaction. 
The energy calibration was conducted using the energy reference from NNDC. The excitation energy of the level at 12.5~MeV was adapted to be 12.5407(2)~MeV from the residual analysis.
The calibration function has a non-linearity at around 1.5 MeV. 
By correcting the non-linearity, the energy accuracy is achieved at 0.3 keV for the overall energy region. 
The photo-peak efficiency in the wide energy range was also measured. 
The extrapolation of the efficiency curve in the low-energy region overestimates the efficiency measured in the high-energy region.
The evaluation of the non-linearity of the energy calibration and photo-peak efficiencies by the actual measurement is crucial 
for accurate spectroscopy of high-energy photons.

\section*{Acknowledgment}
The experiment was performed at the Pelletron facility (joint-use equipment) at the Wako Campus, RIKEN.
The authors thank Prof. Wakasa and Dr. Nishibata at Center for Accelerator and Beam Applied Science (CABAS), Kyushu University for providing the germanium detector (GMX80). This work was supported by Japan Society for the Promotion of Science (JSPS) KAKENHI Grant Number 18H03739 and 19H01357. R. M. is supported by the Forefront Physics and Mathematics Program to Drive Transformation (FoPM), a World-leading Innovative Graduate Study (WINGS) Program, and JSR
Fellowship at the University of Tokyo.

%

\vspace{0.2cm}
\noindent
\let\doi\relax

\end{document}